\documentclass[reprint,
 amsmath,amssymb,
 aps,
pra,
floatfix,
]{revtex4-2}

\usepackage{graphicx}%
\usepackage{dcolumn}%
\usepackage{bm}%
\usepackage{hyperref}%

\usepackage{braket}
\usepackage{xcolor}
\usepackage{multirow}
\usepackage{makecell}
\usepackage{amsmath}
\usepackage{changes}
\usepackage{siunitx}%

\newcommand{\gtwo}{\ensuremath{g^{(2)}(0)}}
\newcommand{\mean}[1]{\langle #1 \rangle}

\newcommand{\up}[1]{\ensuremath{\bar{#1}}}
\newcommand{\low}[1]{\ensuremath{\underline{#1}}}

\newcommand{\etax}[1]{\ensuremath{\eta_{\text{#1}}}}

\begin{document}
\title{Decoy-state quantum key distribution over 227~km with a frequency-converted telecom single-photon source}

\author{Frederik Brooke Barnes}
\author{Christopher L. Morrison}
\author{Zhe Xian Koong}
\author{\\Joseph Ho}
\author{Francesco Graffitti}
\author{Brian D. Gerardot}
\author{Alessandro Fedrizzi}
 \email{a.fedrizzi@hw.ac.uk}
\affiliation{
Institute of Photonics and Quantum Sciences, School of Engineering and Physical Sciences, Heriot-Watt University, Edinburgh EH14 4AS, UK}

\author{Roberto G. Pousa}
\author{John Jeffers}
\author{Daniel K. L. Oi} 
\affiliation{SUPA Department of Physics, University of Strathclyde, Glasgow G4 0NG, UK}

\date{\today}

\begin{abstract}
We implement a decoy-state quantum key distribution scheme using a telecom C-band single-emitter source.
The decoy states are created by varying the optical excitation of the quantum emitter to modulate the photon number distribution.
We provide an analysis of our scheme based on existing security proofs, allowing the calculation of secret key rates including finite key effects. 
This enables us to demonstrate, with a realistic single-photon source, positive secret key rates using our scheme over 227~km of optical fiber, equivalent to a loss tolerance one order of magnitude greater than non-decoy schemes.
This work broadens the scope of single-photon sources in future quantum networks by enabling long-distance QKD with realistic levels of single-photon purity.
\end{abstract}

\maketitle

\section{\label{sec:introduction}Introduction}

Since its inception over forty years ago~\cite{bennett_quantum_2014}, quantum key distribution (QKD) has developed into a mature technology~\cite{xu_secure_2020,pirandola_advances_2020}.
Largely, adoption of weak coherent pulse (WCP) based schemes that exploit simpler hardware, e.g., attenuated diode lasers, and the advent of secure decoy-state protocols~\cite{lo_decoy_2005} have enabled implementations of high-rate and long-distance QKD~\cite{qiu_quantum_2014}.
Secure key rates exceeding 100~Mb/s have been achieved in decoy state protocols~\cite{lo_decoy_2005, wang_beating_2005, li_high-rate_2023,grunenfelder_fast_2023}, while more recent twin-field schemes \cite{lucamarini_overcoming_2018} has demonstrated QKD across 1000~km of fibre~\cite{liu_experimental_2023}, inspiring the prospect of QKD networks at the metropolitan and national scale using trusted nodes \cite{martin_madqci_2024,chen_integrated_2021}.
However, there are constraints on WCP-based QKD, notably coherent states of light contain multi-photon terms which must be dealt with by the decoy state protocol, at the cost of limited key rates and distances compared to an ideal single-photon source (SPS)~\cite{lo_decoy_2005, wang_beating_2005,lucamarini_overcoming_2018}. 
As a secondary effect, the mean photon number per pulse must be kept low---limiting the peak secure transmission rate of WCP-based QKD at short distances. 

Genuine SPSs are therefore an important milestone to surpass repeaterless bounds on secure quantum communication and develop quantum networks toward the vision of a quantum internet~\cite{kimble_quantum_2008, wehner_quantum_2018}.
Recent efforts have demonstrated a trust-free quantum relay for QKD ~\cite{zou2025relayWithDot}, and demonstration of quantum teleportation between dissimilar quantum dots \cite{strobel_telecom_wavelength_2025, laneve_quantum_2025}.
Ideally, SPSs will be developed to be compatible with repeater protocols based on quantum memories~\cite{van_loock_extending_2020,lu_quantum-dot_2021} to move beyond current trusted-node networks.
In the near term, adoption of suitable SPSs can deliver higher rates and longer distance operation of QKD beyond the limits of WCP-based schemes.

Significant progress has been made in the use of single-photon emitters for QKD~\cite{vajner_quantum_2022, bozzio_enhancing_2022, heindel_quantum_2023}.
Sources using open cavity designs have demonstrated high collection efficiencies and mean photon numbers\cite{tomm_bright_2021, ding_high-efficiency_2025}, enabling fractional key rates exceeding the bound for weak coherent pulses~\cite{zhang_experimental_2024}. 
Meanwhile, Purcell-enhanced sources with excited state lifetimes on the order of 100~ps, are capable of operating at GHz clock rates \cite{rickert_high_2024} and have been used in high-rate deployed QKD systems~\cite{yang_high-rate_2023}.
Recent work has begun exploring the development of room-temperature SPSs based on hexagonal boron nitride\cite{samaner2022free,tapcsin2025secure} which face challenges in optimising for brightness and noise.
Beyond prepare-and-measure schemes, high-fidelity entangled states have been produced by single-photon emitters and distributed over metropolitan fibre~\cite{schimpf_quantum_2021, strobel_high-fidelity_2024}, and free-space QKD links~\cite{basso_basset_quantum_2021}.
Additionally, the photon-number coherence has been proposed as another property of light, produced by single-photon emitters, which can be manipulated to enable protocols such as twin-field QKD, or minimised for the consideration of security in prepare-and-measure schemes~\cite{karli_controlling_2024}.
Recently, single-photon emitters frequency converted to telecom wavelengths have enabled QKD demonstrations in optical fibres exceeding 175~km in the lab~\cite{morrison_single-emitter_2023}, and 100~km of deployed fibre~\cite{zahidy_quantum_2024}.
However, there remain outstanding challenges to further increase attainable distances: detector efficiencies; dark count rates; and multi-photon content in the emission to name a few~\cite{pousa_comparison_2024}.

Here we show an increase in the achievable distance of QKD using non-ideal single-photon emitters by adopting techniques from decoy-state protocols, in which states of different mean photon numbers are injected at random to thwart photon-number splitting (PNS) attacks~\cite{lo_decoy_2005}.
We demonstrate this using an InGaAs quantum dot (QD) which produces single photons which are frequency converted into the telecom C-band.
We encode decoy states using a dynamic optical excitation scheme, wherein the pump pulse area is chosen to prepare two target photon number distributions.
We present an updated security analysis for a decoy-state QKD protocol that enables the use of sub-Poissonian light sources, e.g., quantum dots.
Using this updated security model, we demonstrate in the asymptotic limit a non-zero key rate is achievable for 227~km of optical fibre, which is equivalent to 43.4~dB of loss.
Furthermore we present the finite key analysis, using the multiplicative Chernoff bound which has been developed for single-emitter sources in Ref.~\cite{morrison_single-emitter_2023}, to firmly establish the regimes where this new protocol outperforms the traditional, non-decoy protocols.

\section{Results}

We first outline the steps for ensuring security using the decoy-state protocol using a QD-based SPS.
We consider the 2-decoy-state method~\cite{lo_decoy_2005} which requires the sender (Alice) to randomly prepare three states obeying photon number probability distributions labelled as $\delta^{(j)}=\{p_{n | \delta(j)}\}_{n=0}^{\infty}$, where $j\in\{0,1,2\}$, corresponding to the vacuum, signal and decoy states respectively.
The distributions $\delta^{(j)}$ must satisfy, 
\begin{equation} \label{eq:mean-requirement}
    \mean{n_1} > \mean{n_0} + \mean{n_2}
    \quad\mathrm{and}\quad 
    \mean{n_2}>\mean{n_0} \ge 0,
\end{equation}
where $\mean{n_j}$ is the mean photon number of distribution $\delta^{(j)}$. 
To prevent an adversary from discriminating decoys from signals the state preparation must be identical in all degrees of freedom, except the photon number distribution, which is typically achieved by attenuating the encoded single-photon states.
After Alice has sent the states to Bob, they perform reconciliation of the rounds over a public channel.
Bob uses the measurement statistics of the detection and error rates for each distribution to establish lower-bounds on the vacuum $\underline{s}_0^Z$ and single photon $\underline{s}_1^Z$ yields as well as an upper-bound on phase error ($\overline{\phi}_1^Z$) in order to estimate the secret key length~\cite{yin_tight_2020-1,lim_concise_2014},
    \begin{equation} \label{eq:skl}
    \begin{split}
        \ell = \underline{s}_0^Z &+ \underline{s}_1^Z \left[ 1 - h\left( \bar{\phi}_1^Z\right)\right] - f_{EC}n^Zh(E^Z) \\
        & - \log_2\frac{2}{\varepsilon_{\textrm{corr}}} - 6 \log_2 \frac{21}{\varepsilon_\textrm{sec}},
    \end{split}
\end{equation}
where $h(x)$ is the binary entropy function of $x$, $f_{EC}$ is the error correction efficiency, $E^Z$ is the error rate measured in the Z basis, $\overline{\phi}_1^Z$ is the phase error which is measured in the X basis, $\varepsilon_{\textrm{corr}}$ is the correctness parameter, and $\varepsilon_\textrm{sec}$ is the security parameter.

Standard decoy-state implementations use weak coherent pulses which simplify the process of bounding multi-photon rates for distributions of different intensities, as the photon number distributions obey Poisson statistics, typically verified by characterising the source \gtwo$=1$.
In contrast, the photon-number distribution of single-photon emitters depend on the physical properties of the system and must be characterised.
The multi-photon terms in the Fock space can be approximately upper-bounded using the \gtwo\ and $\mean{n}$, as discussed in \cite{morrison_single-emitter_2023}.
The multi-photon fraction produced by a single-emitter is in general smaller than a coherent state with the same $\mean{n}$, and ensuring
\begin{equation} \label{eq:multi-requirement}
    \delta_{n>1}^{(1)} > \delta_{n>1}^{(2)} >\delta_{n>1}^{(0)},
\end{equation}
the same decoy-state parameter estimation method used for WCPs can be used with SPSs (see Appendix~\ref{appendix:QKD_rate_calc} for further details).
 
\begin{figure}[ht!]
    \centering
    \includegraphics{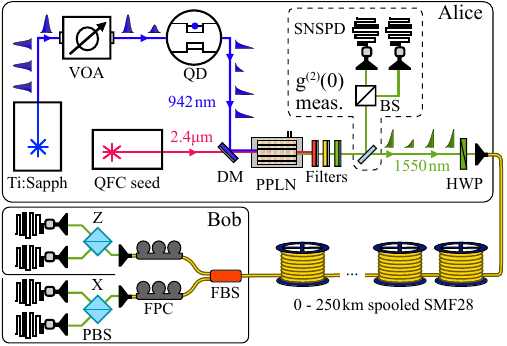}
    \caption{QKD Experimental Setup. 
    A Ti:Sapphire laser generates optical excitation pulses with a repetition rate of 80~MHz and the excitation pulse area is controlled by a VOA.
    The QD is resonantly excited producing single photons at 942~nm that are converted to 1550~nm in a PPLN waveguide using a CW seed laser at 2.4~$\mu$m.
    The 1550~nm photons are extracted using shortpass, longpass and bandpass spectral filters (SP2050, LP1400, BP1550).
    A half-wave plate (HWP) prepares polarisation-encoded photons before coupling into single mode (SMF28) fibre and sent to the receiver.
    A passive receiver performs BB84 measurements using an in-fibre 50:50 beamsplitter (FBS) with fibre polarisation controllers (FPC) and polarising beamsplitters (PBS) to project into X and Z bases, before detection with superconducting nanowire single-photon detectors (SNSPDs).}
    \label{fig:setup}
\end{figure}

We demonstrate this protocol using the setup shown in Figure \ref{fig:setup}.
We use a fiber-coupled electronic variable optical attenuator to modulate the intensity of laser light and thus the pulse area.
The QD, with a ground state emission at T = 4K of $\lambda$ = 942~nm, is driven resonantly with linear polarization and the emission is collected using a cross-polarised, dark-field confocal microscope setup (for more details see Appendix A).
The 942~nm photons are coupled into a 40~mm periodically-poled lithium-niobate (PPLN) ridge waveguide along with a $2.4~\mu$m continuous-wave (CW) laser for difference frequency generation of 942~nm photons to 1550~nm~\cite{morrison_bright_2021}.
The \gtwo\ measurements are taken for the frequency converted photons using a Hanbury-Brown and Twiss setup, which is removed when implementing the QKD scheme.

\begin{figure}[ht!]
\centering
\includegraphics[width=1\columnwidth]{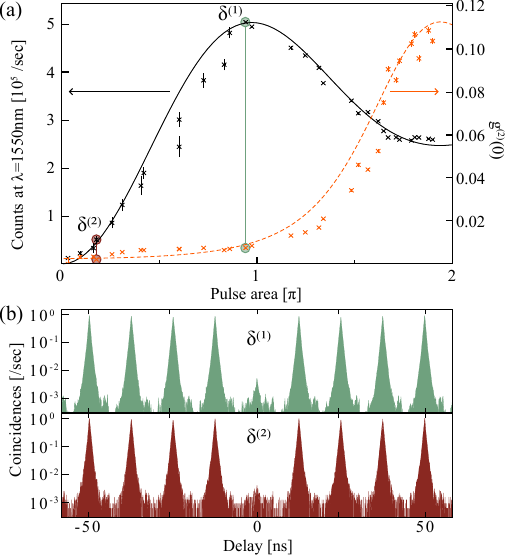}
\caption{Source characterisation.
(a) The measured detection count rates (black data, left-hand axis)  and \gtwo\ (orange data, right-hand axis) of the resonantly excited QD emission after QFC from 942~nm  to 1550~nm as the excitation pulse area is varied from $[0, 2\pi]$.
Solid lines are generated from a simplified two-level model, see Appendix~\ref{sec:qd-model} for details. Error bars are estimated assuming Poissonian statistics.
(b) Coincidence measurements for the signal state ($\delta^{(1)}$) with \gtwo\ of $0.0159\pm 0.0002$ and decoy state ($\delta^{(2)}$) with \gtwo\ of $0.0155\pm 0.0005$.
}
\label{fig:characterisation}
\end{figure}

We characterised the source of 1550~nm photons by measuring the count rates and \gtwo\ as a function of excitation pulse area, as shown in Figure~\ref{fig:characterisation}(a).
The \gtwo\ increases with the pulse area, indicating the re-emission due to the QD emitting before the end of the excitation pulse ~\cite{dada_indistinguishable_2016}, see Appendix~\ref{sec:qd-model} for details on modelling this behaviour.
From this, we select two pumping regimes for preparing the signal ($\delta^{(1)}$) and decoy ($\delta^{(2)}$) states as indicated on the plots.
The signal state is chosen to maximise the mean photon number, while the decoy state was chosen to optimise the performance of the decoy-state protocol.
Notably the multi-photon components of the distributions obey the relationship in Equations~\ref{eq:mean-requirement}.
For $\delta_{n,1}$ we used a pulse-area of $0.96~\pi$ resulting in $\sim5\times10^5$ counts per second (cps) wit $\mean{n_1}=\num{1.16e-2}$; $\delta_{n,2}$ used $0.18~\pi$ which produced $\sim5\times10^4$ cps with $\mean{n_2}=\num{7.6e-3}$; and we set $\delta_{n,3}$ as vacuum with $\mean{n_3}=0$ (equivalent to no driving). 

Alice prepares one of four polarisation-encoded states (H,V,A,D) using a HWP, for each $\delta^{(j)}$.
While we consider the efficient choice of key-generation basis probability, $p_Z$, for assessing the performance of the protocol, we implement a passive preparation and measurement scheme for convenience in this experiment. 
Alice then sends the states over variable lengths of single-mode fibre to the receiver.
Bob projects the states into the X or Z basis using a pair of fibre polarisation controllers and polarising beam splitters, and records the detection events with four single-photon detectors as shown in Figure \ref{fig:setup}. Additionally, Bob uses a time-gating window of 1.25~ns to mitigate the effect of dark counts. 

From the measurement statistics we perform parameter estimation and establish the secure key rates.
The single-photon decoy method differs from the coherent-state decoy method in that we use new bounds for $\bar{\phi}$, $s_0^Z$, and $s_1^Z$. 
For $\phi_Z$, the experimental parameters required at this stage are the number of detection events in the $X$ basis for each distribution, $n_{X,\delta^{(j)}}$, and the number of errors, $m_{X,\delta^{(j)}}$, for a given block length.
We estimate these through measurement of the detection and error events per pulse for each distribution, $p_{\text{c}, \delta^{(j)}}$ and $p_{\text{err}, \delta^{(j)}}$ respectively.
These are measured by analysing Bob's count rates with respect to the repetition rate, and the ratio of events where Bob measures the incorrect state compared to that prepared by Alice given that Bob chooses the correct basis.
This allows extrapolation of $n_{Z,\delta^{(j)}}$ and $m_{Z,\delta^{(j)}}$ for different raw key block lengths.
Similarly, $\underline{s}_0^Z$ and $\underline{s}_1^Z$ can be derived from $n_{Z,\delta^{(j)}}$ and the previously characterised source parameters and user-defined protocol parameters.
A detailed description of the parameter estimation is provided in Appendix~\ref{appendix:QKD_rate_calc}.

\begin{figure}[ht!]
\centering
\includegraphics{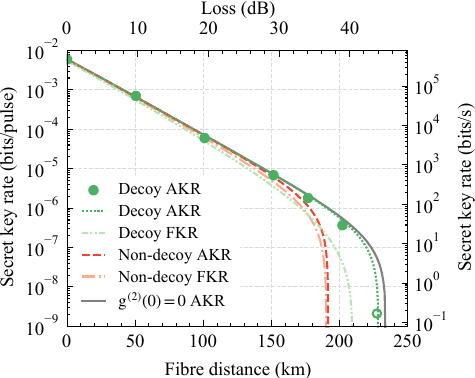}
\caption{Comparison of Non-Decoy and Decoy protocols.
AKR and FKR ($10^{11}$ signals sent) as a function of channel loss for the Non-Decoy BB84 protocol (red) and Decoy protocol (green).
Key rate estimates from the experimental data (filled green circles) follow our system model.
AKR for a source with the same $\langle n \rangle$ as our source and $g^{(2)}(0)=0$ is shown to achieve positive key rates at only slightly greater distances than our source with the Decoy protocol.
The unfilled green circle is the experimentally measured AKR when using a smaller gating window of 250~ps.}
\label{fig:qkd}
\end{figure}

The asymptotic key rate (AKR) at each distance is calculated for the non-decoy protocol as in Ref.~\cite{morrison_single-emitter_2023} and from Equation~\ref{eq:skl} for the Decoy protocol, as shown in Figure \ref{fig:qkd}a.
AKR extracted from the parameters estimated in our experiment are shown to match the expected key rates of our modelled system.
In the low-loss regime, both protocols perform similarly as the ratio of prepared signal states to decoy states approaches unity.
In the high-loss regime, the Decoy protocol shows a clear improvement, allowing positive key rates beyond 227~km (44~dB) compared to 191~km (37~dB) for the standard non-Decoy protocol. 
In this regime, the decoy states $\delta^{(2)}$ and $\delta^{(0)}$ do not themselves contribute to the secure key as their QBER exceeds the Holevo bound for secure key generation, but they enable lower bounds on the single photon yield leading to greater key rates. 
Notably, the Decoy protocol comes close to saturating the Non-Decoy AKR of a perfectly pure SPS ($g^{(2)}(0)=0$) of the same brightness, which we model as achieving positive key rates up to 233~km (45 dB).
This indicates that, in the asymptotic limit, the Decoy protocol is limited by the single-to-noise ratio at Bob rather than the single-photon impurity of Alice's source.
This is evidenced by the need to minimise the effect of dark counts by using a tighter gating window of 250~ps in order to extract a positive key rate from our experimental data at a distance of 227~km.

While the AKR provides an upper bound on the tolerable loss for secure key generation, it is important to characterise the performance of a protocol with finite block sizes that can be generated in realistic timescales.
Figure~\ref{fig:qkd} shows the FKR for each protocol in the scenario that Alice sends $10^{11}$ pulses, equivalent to a practical acquisition time of approximately 20 minutes.
In this scenario, the Non-Decoy protocol slightly outperforms the Decoy protocol for low losses.
The regime of Decoy-protocol advantage beings at 175~km (33~dB), and by 190~km (37~dB) the Non-Decoy FKR drops to zero while the Decoy FKR remains above 10 bits/s.
Ultimately, the Decoy FKR in this scenario remains positive up to 209~km (40~dB).
Although we use tight bounds for parameters of both protocols, including Chernoff bounds for finite sampling statistics ~\cite{tamaki_decoy-state_2016, morrison_bright_2021, yin_tight_2020-1}, we find that the Decoy protocol FKR does not follow the AKR as closely as the Non-Decoy protocol.
We attribute this to the Decoy protocol having a greater number of parameters than the Non-Decoy protocol, leading to a greater fraction of raw key being expended by parameter estimation. 

\section{Discussion}\label{dicussion}
We note recent work that utilizes photon statistics, such as $g^{(2)}(0)$, observed by the receiver to detect the presence of an eavesdropper performing certain PNS attacks~\cite{cholsuk2025decoy}. 
Under certain assumptions, this allows the range and rate of QKD to be extended for sources with high values of $g^{(2)}(0)$ without the use of a decoy state, but is not secure against the most general PNS attack. 
We also note that the work of~\cite{bloom2025decoy} makes a strong assumption that Fock state contributions of three or more photons can be totally neglected for their source.
This assumption is not necessarily valid or secure to assume, hence the applicability of their protocol and analysis is restricted. 
Meanwhile, other work \cite{foletto_security_2022} has focused on particular photon number distributions. 
The assumptions of the Fock state population distribution and decay in our work are less restrictive and are valid for a greater class of SPSs and driving conditions.

This security and composability of our protocol is maintained through existing security analyses.
Additionally, it allows us to employ techniques from the well-established field of WCP QKD to create tight bounds on key rate for asymptotic and finite block lengths. 
In the finite-key regime, the Decoy protocol only confers an advantage for our source at the upper limit of practical acquisition times, but sources with similar Purcell-enhancement have been successfully operated at GHz repetition rates~\cite{rickert_high_2024}. 
Additionally, state-of-the-art QDs have individually exhibited brightness of at least one order of magnitude greater than our QD~\cite{zhang_experimental_2024, tomm_bright_2021} before frequency conversion. 
Thus, we expect the regime where decoy states confer an advantage to shift increasingly towards practical acquisition times, with the crossover point potentially decreasing to the order of 10~s with existing SPSs.
Because the decoy state method is applicable beyond prepare-and-measure QKD, such as measurement-device-independent and twin-field protocols, our methods could also help to enable non-ideal SPSs in more advanced quantum communication protocols.

Our methods could be readily implemented with intensity modulation applied to the QD emission, in a similar fashion to WCP QKD, rather than dynamic optical excitation.
However, our dynamic QD excitation method avoids the insertion loss on the generated single-photon states, which is typically $>2$~dB for an in-fibre Mach-Zehnder (MZ) modulator. While we use a VOA for excitation modulation, an EOM modulator could enable active encoding of decoy states at high repetition rates by pulse-carving~\cite{dada_indistinguishable_2016} or by selective attenuation of a pulsed laser.
In contrast to a WCP-source, the brightness of a SPS cannot be increased to compensate for this additional loss.
Therefore, our method could be an important tool for exploiting photon-number as a useful degree of freedom for single-emitters in QKD~\cite{bozzio_enhancing_2022, karli_controlling_2024}, and further exploration of dynamic excitation methods, such as phonon-assisted excitation, may yield additional advantages such as removing the requirement for discrete phase randomisation~\cite{lo_phase_2005, cao_discrete-phase-randomized_2015}. 
Recent work has shown active encoding of BB84 polarisation states can be achieved using similar sources with sufficiently low encoding error while also being compatible with deployed optical fibre~\cite{zahidy_quantum_2024}.
Therefore, we expect our protocol to be readily implemented in a more realistic setting with established methods.

In summary, we provide security analysis and experimental demonstration, with a quantum-dot based telecom C-band SPS, that decoy states can significantly increase the maximum possible key distribution distance, enabling the distribution of a secret key over 227~km and in the asymptotic regime and maintains an advantage over non-decoy protocols for practical acquisition times at long distances.
Notably, we find that the FKR performance using decoy states for our imperfect SPS, approaches the performance for a SPS with an ideal \gtwo=0, i.e., the best-case scenario.
Furthermore, we expect our method to be readily accessible for utilisation in a deployed fibre network for practical QKD.
Therefore, this work loosens the bound on single-photon purity required to distribute keys over long distances, opening the door to practical QKD for low purity SPSs, and expands the scope of high performance quantum-dot sources as a general-purpose source in future quantum networks.

\begin{acknowledgments}
This work was supported by the EPSRC Quantum Technology Hub in Quantum Communication (EP/T001011/1), International Network in Space Quantum Technologies (EP/W027011/1), and the Integrated Quantum Networks Research Hub (EP/Z533208/1). B.D.G. acknowledges support from the Royal Academy of Engineering for a Chair in Emerging Technology. 
\end{acknowledgments}

\appendix

\section{Single-photon source}
\label{sec:source}
The single-photon source (SPS) is based on an InGaAs QD in a micropillar cavity embedded in a p-i-n diode structure, which allows tuning of the neutral exciton into resonance with the cavity modes via the Stark effect.
The QD is cooled to 4~K using a closed-loop liquid helium cryostat (Attodry).
Excitation pulses are created by a wavelength-tunable Ti:Sapphire laser (Coherent Chameleon) which outputs mode-locked femtosecond pulses. 
This is spectrally shaped using a 4$f$-pulse shaper (APE) to produce transform limited pulses of 0.1~nm bandwidth ($\sim18$~ps pulses).
The excitation power is controlled by an electronic variable optical attenuator (Thorlabs EVOA) before attenuation using a 99/1 beam splitter. The EVOA has a bandwidth of \qty{1}{\kilo\hertz}, while an electro-optic modulator could provide a bandwidth up to the \qty{}{\giga\hertz} regime and enable pulse-by-pulse modulation of the excitation power.
The excitation laser is filtered from the emission at 942.2~nm by cross-polarisation extinction using a combination of HWP, QWP, polarising beamsplitter, and linear polarisers in the excitation and collection, achieving extinction ratios $>10^7$.
The single-photon emission is collected into a single mode fibre (780HP) before being sent to the quantum frequency conversion (QFC) setup.
The QFC setup consists of a periodically-poled lithium niobate ridge waveguide (NTT) held at the relevant phase matching temperature by a temperature-controlled oven (Covesion) for difference-frequency generation of 1550~nm from 942.2~nm, seeded by a 2.4~$\mu$m Cr:ZnSe CW laser (home-built) which is optically pumped by a 1900~nm Thulium fibre laser (IPG).
The polarisation conditions for Type-I conversion are satisfied by co-aligning the single-photons and seed laser using quarter- and half-waveplates with the vertical polarisation supported by the waveguide, and combined using a dichroic mirror, before being coupled into the waveguide using an aspheric lens.
The waveguide output is collimated with an aspheric lens and spectrally filtered using a 2050~nm short pass filter to remove the seed laser as well as 1400~nm low pass and 1550$\pm$3~nm band pass (Alluxa) filters to remove unwanted light from inelastic scattering processes. 
 and methods.

\section{Quantum dot modelling} 
\label{sec:qd-model}
The quantum dot is modelled as a two-level system with a ground state, $\ket{g}$, and excited state $\ket{e}$.
The lowering operator, $\sigma$, is defined as $\ket{g} \bra{e} $.
The Hamiltonian of the QD is $H_{\text{qd}} = \omega_{\text{qd}} \ket{e}\bra{e}$.
The Hamiltonian of the QD interaction with a CW-field is $H_{\text{cw}}= \frac{\Omega}{2} (\sigma + \sigma^{\dagger})$, where $\Omega$ is the Rabi-frequency.
For a laser with a Gaussian pulse-profile, $\Omega = \frac{A}{\sqrt{\tau_p^2  \pi}} e^{-(t-t0) / \tau_p)^2}$, where $A$ is the pulse area such that for $A=n\pi$ the quantum dot will undergo $\frac{n}
{2}$ Rabi cycles.
The total Hamiltonian of the system is $H = H_{\text{qd}} + H_{\text{cw}}$ \cite{nazir_photon_2008}.

Interactions with the environment are described using collapse operators.
Dephasing due to phonon processes is described by $\sqrt{B  d^2}  \sigma^{\dagger} \sigma$, where $B$ is the phonon parameter and $d$ is the dipole strength.
The process of spontaneous emission is described by $\sqrt{\Gamma}\sigma$ \cite{fischer_dynamical_2016}. 

The dynamics of the system, including the processes represented by the collapse operators $a_n$, are included as Louivillian superoperators within the quantum-optical master equation:

\begin{equation}
    \frac{d\rho}{dt} = -\frac{i}{\hbar}[H, \rho] + \sum_n \frac{1}{2}[2C_n\rho C_n^{\dagger} - \rho C_n^{\dagger}C_n - C_n^{\dagger}C_n\rho] = \mathcal{L}\rho,
\end{equation}

where $C_n = \sqrt{\gamma_n}a_n$. The model was implemented using QuTiP \citep{johansson_qutip_2013}, employing the formalism for using the quantum regression theorem with time-dependent Loiuvillians \citep{fischer_dynamical_2016}. This allows the calculation of correlations of the form
\begin{equation}
    G(t,\tau) = \langle A(t)B(t,\tau)C(t)\rangle.
\end{equation}
For instance, $G^{(2)}$ can be calculated by setting $A(t)=a^\dagger(t)$, $B(t)=a^\dagger(t+\tau)a(t+\tau)$, and $C(t)=a(t)$. This can then be normalised by the expected photon number to obtain \gtwo.

\section{Decoy protocol secret-key length}
\label{appendix:QKD_rate_calc}
In this section, we provide bounds for the parameters used to estimate the secret-key length, as in Equation \ref{eq:skl}, from the parameters estimated during the protocol and characterisation of the SPS. 

To estimate ${n}_{X,1}$, we adapt the method in \cite{lim_concise_2014} and write

\begin{multline}
    \frac{p_{0|\delta(b)} n_{X,\delta(a)}}{p_{\delta(a)}} - \frac{p_{0|\delta(a)} n_{X,\delta(b)}}{p_{\delta(b)}} = \\\frac{n_{X,1}}{p_1} \bigg(p_{0|\delta(b)}p_{1|\delta(a)} - p_{0|\delta(a)}p_{1|\delta(b)} \bigg) + \\\sum_{n=2}^{\infty}\frac{n_{X,n}}{p_n} \bigg(p_{0|\delta(b)}p_{n|\delta(a)} - p_{0|\delta(a)}p_{n|\delta(b)} \bigg),
\end{multline}
where $n_{X,\delta{(j)}}$ are the sifted events in the $X$-basis from each distribution, $n_{X,n}$ are the sifted events attributable to $n$-photons, and $p_n = \sum_jp_{\delta(j)}{p_{n|\delta(j)}}$ is the probability that Alice sends $n$-photons across all prepared states.
By solving for $n_{X,1}$, and introducing a third distribution, $\delta(c)$, we may write the bound
\begin{multline}
     n_{X,1} \geq \frac{p_1}{p_{0|\delta(b)} p_{1|\delta(a)} - p_{0|\delta(a)} p_{1|\delta(b)}} \bigg(\frac{p_{0|\delta(b)n_{X,\delta(a)}}}{p_{\delta(a)}} \\- \frac{p_{0|\delta(a)n_{X,\delta(b)}}}{p_{\delta(b)}} - p_{0|\delta(b)} \sum_{n=2}^{\infty} \frac{p_{n|\delta(c)}n_{X,n}}{p_n} \bigg),
 \end{multline}
 on the condition that ${p_{n|\delta(c)}} \geq {p_{n|\delta(a)}}$ for $n\geq 2$, which may be satisfied by setting the signal as $\delta(c)=\delta^{(1)}$, and the decoy as $\delta(a)=\delta^{(2)}$. Again, solving for $n_{X,1}$ and taking the vacuum state as $\delta(b)=\delta^{(0)}$, we find
\begin{multline}
    n_{X,1} \geq \frac{\underline{p}_1}{\overline{p}_{1 | \delta(c)} - \underline{p}_{1 | \delta(a)} } \bigg( \frac{n_{X,\delta(c)}}{p_{\delta(c)}} - \frac{n_{X,\delta(a)}}{p_{\delta(a)}} \\
    - \overline{p}_{0 | \delta(c)} \frac{\overline{n}_{X,0}}{\underline{p}_0} + \underline{p}_{0 | \delta(a)} \frac{n_{X,\delta(0)}}{p_{\delta^{(0)}}} \bigg),
\end{multline}
where we have replaced each parameter with its relevant upper or lower bound.

To estimate $\up{n}_{X,0}$, we note that $n_{X,\delta(0)} = \frac{p_{\delta(0)}}{p_0}n_{X,0}$ and find 
\begin{equation}
    \up{n}_{X,0} = n_{X,\delta(0)} \left(1 + \frac{p_{\delta(a)}}{p_{\delta(0)}} \up{p}_{0|\delta(a)} + \frac{p_{\delta(c)}}{{p_{\delta(0)}}} \up{p}_{0|\delta(c)} \right),
\end{equation}
using the definition of $p_n$ given above.

We upper-bound the single-photon phase error as $\up{\phi^X_1} = \frac{\up{m}_{Z,1}}{\low{n}_{Z,1}}$. 
To estimate ${m}_{Z,1}$, the number of sifted single-photon errors in the $Z$ basis, we consider the difference in the number of errors due to two distributions as
\begin{equation} \label{eq:diff mx}
      m_{Z,\delta(c)} - m_{Z,\delta(b)} = \sum_{n=0}^{\infty}\frac{m_{Z,n}}{p_n}(p_c p_{n|c} - p_b p_{n|b}).
\end{equation}
By choosing $\delta(c)=\delta(1)$ and $\delta(b)=\delta(0)$, we can ensure $p_c p_{n|c} \geq p_b p_{n|b}$ for $n\neq1$ by choosing $p_c$ and $p_b$ such that $p_c p_{0|c} \geq p_b$. 
Hence, by re-arranging Equation \ref{eq:diff mx}, we may write
\begin{equation}
      \up{m}_{X,1} = \frac{\up{p}_1 (m_{X,\delta(c)} - m_{X,\delta(b)})}{p_c \low{p}_{1|c} - p_b \up{p}_{1|b}},
\end{equation}
where we have replaced each photon-number probability with its relevant bound.

Our treatment of finite-key effects follows the statistical fluctuation analysis of \cite{yin_tight_2020-1}, in which the relevant Chernoff bound (or its variant) is used to bound estimates of the expected number of detection events from the observed number of events (and vice versa). Similarly, the expected number of errors in the key-generation basis are estimated from the observed events in the parameter-estimation basis, using the results in \cite{yin_tight_2020-1} based on treating the problem as a random sampling without replacement problem as in \cite{korzh_provably_2015}.

\section{Bounding photon-number probabilities}
To bound $p_{0|\delta(j)}$ and $p_{1|\delta(j)}$, as required in Appendix \ref{appendix:QKD_rate_calc}, we note that $p_{0|\delta(j)} + p_{1|\delta(j)} + p_{m|\delta(j)} = 1$ where $p_{m|\delta(j)} = \sum_{n=2}{p_{m|\delta(j)}}$ is the multi-photon content of the distribution. 
The upper-bound on the multi-photon content, $\up{p}_{m|\delta(j)}$, is given by $\up{p}_{2|\delta(j)} = \frac{\gtwo_{\delta(j)} {\mean{n}^2_{\delta(j)}}}{2} \cite{waks2002}$, and  $\low{p}_{m|\delta(j)}=0$. We take $\up{p}_{1|\delta(j)} = \mean{n}_{\delta(j)}$ by noting $\mean{n} = p_{1|\delta(j)} + \sum_{n=2}^{\infty}p_nn \geq p_{1|\delta(j)}$. 
By taking the definitions of $\mean{n}$ and $\gtwo = \sum_{n=0}^{\infty}\frac{n(n-1_)p_n}{\mean{n}^2}$, we may write $\low{p}_{1|\delta(j)} = \mean{n} - \gtwo \mean{n}^2 = p_{1|\delta(j)} + \sum_{n=2}^{\infty}2n-n^2 \leq p_{1|\delta(j)}$. Finally, we find bounds for the vacuum probability as $\low{p}_{0|\delta(j)} = 1 - \up{p}_{m|\delta(j)} - \up{p}_{1|\delta(j)}$, and $\up{p}_{0|\delta(j)} = 1 - \low{p}_{m|\delta(j)} - \low{p}_{1|\delta(j)}$. Therefore, all required bounds for photon-number probabilities may be derived from measurement of $\mean{n}$ and $\gtwo$. 

\section{Modelling secret key lengths}
To model the QKD system's performance we estimate the number of sifted events in the $X$-basis for each distribution, $n_{X,\delta(j)}$, and the number of errors, $m_{X,\delta(j)}$.
For the detection events, we write $n_{X,\delta(j)} = N_S p_{\delta(j)}p_{X}^2 p_{\text{click}, \delta(j)}^{X} $, where $N_S=Rt_{\text{meas}}$ is the number of states sent by Alice, $R$ is the repetition rate, and $p_{\text{click}, \delta(j)}^{X}$ is the per-pulse probability Bob registers a click given Alice and Bob choose the $X$-basis and Alice chooses to send a state with photon-number distribution $\delta(j)$. 
Similarly, for the errors, we write $m_X = N_Sp_X^2 p_{e, \delta{(j)}}^X$, where $p_{e, \delta{(j)}}^X$ is the per-pulse probability that Bob records an error in the $X$-basis. 
The same can be calculated for the $Z$-basis. 
We estimate $p_{\text{click}, \delta(j)}^{X}$ and $p_{e, \delta{(j)}}^X$ with a model of the QKD system or derive them from experimental measurements.

To model $p_{\text{click}, \delta(j)}^{X,Z}$, we use Equation 4 from \cite{morrison_single-emitter_2023} applied it to each $\delta(j)$. 
To model $p^{X,Z}_e$, we use 
\begin{equation} \label{eq: p_e}
p_{e, \delta(j)} = c_{\text{dt}}\sum_{n=0}^{\infty} \left[\frac{p_{dc}}{2} + p_{\text{mis}}(1 - p_{dc})(1 - (1-\eta)^n) \right]p_{n|\delta(j)},
\end{equation}
where $p_{\text{mis}}$ is the error probability representing the overall misalignment of Alice and Bob's state preparation and detection stages, $\eta = \etax{ch} \etax{det}$, $\etax{ch}$ is the channel transmission efficiency, $\etax{det}$ is the detection efficiency, and $c_{\text{dt}}$ is a factor accounting for the dead-time of the detectors, given by 
\begin{equation}
c_{\text{dt}} = \frac{1}{1 + R \tau p_{\text{click}}}, 
\end{equation}
 where $\tau$ is the dead-time of the detector.
 
To estimate these probabilities from our experiment, we note that, for each fibre distance, Alice continuously prepares each polarisation state ($\text{prep}$) over a measurement time $t_{\text{meas}}$. This is repeated for each fibre length with each of the signal and decoy distributions, using $t_{\text{meas}}$ listed in Table \ref{tab:Measurement_times}.
Bob acts as a passive receiver by employing a 50/50 BS, collecting counts on four detectors corresponding to the four polarisation states ($\text{meas}$).
These detections events are generated for each $\delta^{(j)}$, thus yielding count rates denoted as $\text{counts}_{\text{meas}|\text{prep},\delta^{(j)}}$. For instance, when Alice prepares $A$ in the signal state, Bob's counts in the $D$ detector are given by $\text{counts}_{\text{D}|\text{A},\delta(1)}$. Hence, 
\begin{equation}
p_{\text{click}, \delta(j)}^{Z} = \frac{\sum_{\text{meas}=\{A,D\}}\sum_{\text{prep}=\{A,D\}} \text{counts}_{\text{meas}|\text{prep},\delta^{(j)}}}{{t_\text{meas}R}},
\end{equation}
for the $Z$-basis, and may be calculated similarly for $X$-basis with $H$ and $V$. For the error probability when Alice prepares a state, say $D$,
\begin{equation}
   p^D_{e,\delta(j)} = \frac{\text{counts}_{\text{A}|D,\delta(j)}}{\text{counts}_{\text{A}||D,\delta(j)} + \text{counts}_{\text{D}|D,\delta(j)}},
\end{equation}
The error in each basis may then be taken as the average, $p^{Z,X}_{e,\delta(j)} = \frac{1}{2}(p^{D,H}_{e,\delta(j)} + p^{A,V}_{e,\delta(j)})$.
These click and error probabilities are then used to derive the experimental asymptotic key rates in Figure \ref{fig:qkd}.

\begin{table}[]
\centering
\begin{tabular}{|c|c|c|c|}
\hline
\textbf{\makecell{Fibre \\ distance  \\ (km)}} & \textbf{\makecell{Channel \\ loss (dB)}} & \textbf{Distribution} & \textbf{\makecell{Measurement \\ time $t_{\text{meas}}$ (s)}} \\ \hline

\multirow{2}{*}{\makecell{0, 50.4, 100.9 \\ and 151.5}} & \multirow{2}{*}{\makecell{0, 9, 18.2, \\ and 28.3}} & signal & 60 \\ \cline{3-4} & & decoy & 60 \\ \hline

\multirow{2}{*}{176.7} & \multirow{2}{*}{33.4} & signal & 120 \\ \cline{3-4} & & decoy & 120 \\ \hline

\multirow{2}{*}{202.0} & \multirow{2}{*}{38.3} & signal & 120 \\ \cline{3-4} & & decoy & 240 \\ \hline

\multirow{2}{*}{227.3} & \multirow{2}{*}{43.4} & signal & 240 \\ \cline{3-4} & & decoy & 360 \\ \hline
\end{tabular}
\caption{Measurement times for each distribution and optical fibre distance (channel loss).}
\label{tab:Measurement_times}
\end{table}

\end{document}